\documentclass[aps,prl,preprint]{revtex4}
\usepackage{bbm,slashed}
\usepackage{color}
\usepackage{graphicx,pstricks}
\usepackage{times}
\usepackage{psfrag}
\definecolor{Yellow}{cmyk}{0,0,1,0}
\definecolor{Dandelion}{cmyk}{0,0.29,0.84,0}
\definecolor{SpringGreen}{cmyk}{0.26,0,0.76,0}
\definecolor{GreenYellow}{cmyk}{0.15,0,0.69,0}
\definecolor{Goldenrod}{cmyk}{0,0.1,0.84,0}
\definecolor{RoyalBlue}{cmyk}{1,0.5,0,0}
\definecolor{shadecolor}{gray}{.80}
\definecolor{wred}{rgb}{0,.8,0}
\definecolor{wbl}{rgb}{.3,.3,1}
\definecolor{wpurple}{rgb}{.8,0,.7}

\newcommand{\wblue}{\color{wbl}}
\newcommand{\wpurple}{\color{wpurple}}

\font\af=msbm12
\font\aff=eurm10
\newcommand{\Nt}{N_\mathrm{t}}
\newcommand{\Ns}{N_\mathrm{s}}
\newcommand{\Vs}{V_\mathrm{s}}
\newcommand{\Nc}{N_\mathrm{c}}
 \DeclareMathAlphabet{\boldmathe}{T1}{cmr}{bx}{it}

\newcommand{\mbf}[1]{\boldmathe{#1}}
\newlength{\graphwidth}
\newlength{\subgraphwidth}
\newcommand{\refs}[1]{(\ref{#1})}

\newcommand{\eqnl}[2]{\begin{eqnarray}#1
\label{#2}\end{eqnarray}}
\begin{document}
\def\lamz{{^{z\!}\lambda}}
\def\lamzk{{^{z_k\!}\lambda}}
\def\diz{{^{z}\mathcal D}}
\def\dizk{{^{z_k}\mathcal D}}
\def\cC{{\mathcal C}}
\def\cD{{\mathcal D}}
\def\cE{{\mathcal E}}
\def\cF{{\mathcal F}}
\def\cG{{\mathcal G}}
\def\cP{{\mathcal P}}
\def\cS{{\mathcal S}}
\def\cW{{\mathcal W}}
\def\cZ{{\mathcal Z}}
\def\Zs{\mbox{\aff Z}}
\def\Z{\mbox{\af Z}}
\def\ring{\mathaccent"7017}
\def\id{\mathbbm{1}}
\def\mtxt#1{\quad\hbox{{#1}}\quad}
\def\mbx{\mbf{x}}
\def\mbv{\mbf{v}}
\def\tr{{\rm tr}}
\def\di{\slashed{D}}
\def\lam{\lambda}
\title{Spectral sums of the Dirac-Wilson Operator and their
relation to the Polyakov loop}

\author{Franziska Synatschke, Andreas Wipf and Christian Wozar}
\affiliation{Theoretisch-Physikalisches Institut,
Friedrich-Schiller-Universit{\"a}t Jena, Max-Wien-Platz 1, 07743
Jena, Germany}
  
\begin{abstract}
We investigate and compute spectral sums of the Wilson lattice
Dirac operator for quenched $SU(3)$ gauge theory. It is demonstrated 
that there exist sums which serve as order parameters for the 
confinement-deconfinement phase transition and get their main 
contribution from the IR end of the spectrum. They
are approximately proportional to the Polyakov loop.   
In contrast to earlier studied spectral sums some of them
are expected to have a well-defined continuum limit.
\end{abstract}
\maketitle

\section{Introduction}
Confinement and chiral symmetry breaking are prominent
features of strongly coupled gauge theories.
If the gauge group contains a non-trivial center $\cZ$, then the 
traced Polyakov loop
\cite{Polyakov:1978vu,Susskind:1979up}
\eqnl{
L_\mbx=\tr_c\, \cP_\mbx,\quad \cP_\mbx=
\prod_{\tau=1}^{\Nt} U_0(\tau,\mbx)}{polya1}
serves as an order parameter for confinement in pure gauge theories
or (supersymmetric) gauge theories with matter in the adjoint representation.
The dynamics of $L_\mbx$ near the phase transition
point is effectively described by generalized Potts models \cite{Yaffe:1982qf,Wozar:2006fi}.
Here we consider the space-independent expectation values 
$\langle L_\mbx\rangle$ only and thus may replace $L_\mbx$ by 
its spatial average
\eqnl{
L=\frac{1}{V_s}\sum_\mbx L_\mbx,\quad  \Vs=\Ns^{d-1}.}{polya2} 
The expectation value $\langle L\rangle$
is zero in the center-symmetric confining phase and
non-zero in the center-asymmetric deconfining phase.

Chiral symmetry breaking, on the other hand, is related to an unusual distribution
of the low lying eigenvalues of the Euclidean Dirac operator $\cD$ \cite{Leutwyler:1992yt}. In the 
chirally broken low-temperature phase the typical distribution is dramatically different 
from that of the free Dirac operator since a typical level density $\rho(\lam)$ 
for the eigenvalues per volume does not vanish for $\lam\to 0$. 
Indeed, according to the celebrated Banks-Casher relation \cite{Banks:1979yr}, 
the mean density in the infrared is proportional to the quark condensate, 
\eqnl{
\langle \rho(0)\rangle=-\frac{1}{\pi}
\langle 0\vert \bar\psi\psi\vert 0\rangle.}{banks1}
Which class of gauge field configurations gives rise to this unusual
spectral behavior has not been fully clarified. It may be
a liquid of instanton-type configuration \cite{Schafer:1996wv}.
Simulations of finite temperature $SU(3)$ gauge theory without dynamical quarks
reveal a first order confinement-deconfinement  phase transition
at 260 MeV. At the same temperature the chiral condensate vanishes. This 
indicates that chiral symmetry breaking and confinement are most
likely two sides of a coin (\cite{Kogut:1982rt}, for a review see e.g. 
\cite{Karsch:2001cy}). 
  
Although it is commonly believed that confinement and chiral 
symmetry breaking are deeply related, no analytical evidence of such
a link existed up to a recent observation by Christof 
Gattringer \cite{Gattringer:2006ci}. His formula holds for lattice regulated
gauge theories and is most simply stated for Dirac operators with nearest neighbor 
interactions. Here we consider fermions with ultra-local
and $\gamma_5$-hermitean Wilson-Dirac operator
\eqnl{
\langle y\vert \cD\vert x\rangle=(m+d)\delta_{xy}
-\frac{1}{2}\sum_{\mu=0}^{d-1} \left((1+\gamma^\mu)U_{-\mu}(x)\delta_{y,x-e_\mu}
+(1-\gamma^\mu)U_\mu(x)\delta_{y,x+e_\mu}\right),}{DWOperator}
where $U_{\pm\mu}(x)$ denotes the parallel transporter from
site $x$ to its neighboring site $x\pm e_\mu$ such that $U_{-\mu}(x+e_\mu)U_\mu(x)=\id$ 
holds true. 
Since we are interested in the finite temperature behavior we choose
an asymmetric lattice with $\Nt$ sites in the temporal direction and 
$\Ns\gg \Nt$ sites in each of the $d-1$ spatial directions. We impose periodic
boundary conditions in all directions. The 
\eqnl{ 
{\rm dim}(\cD)=V\times 2^{[d/2]}\times N_c,\quad V=\Nt\times \Vs,}{hilfs1}
eigenvalues of the Dirac operator in a background field
$\{U_\mu(x)\}$ are denoted by $\lam_p$. The non-real ones
occur in complex conjugated pairs since $\cD$ is $\gamma_5$-hermitian.
If $\Nt$ and $\Ns$ are both even, then $\lam_p\to 2(d+m)-\lam_p$ is a 
further symmetry of the spectrum.\\
\begin{minipage}[t]{6.5cm}
\psset{unit=1.1cm}
\begin{pspicture}(-0.3,0.5)(6.7,6.5)
\multirput(0.5,1)(1,0){6}{\psline(0,0)(0,5)}
\multirput(0,1.5)(0,1){5}{\psline(0,0)(6,0)}
\multirput(0.5,2.5)(1,0){6}{\psline[linewidth=.8mm,linecolor=red](0,0)(0,1)}
\rput(5.3,2.3){$x$}
\rput(6.2,2.5){$\tau$}
\rput(6.25,3){${\red z\,U_0(x)}$}
\end{pspicture}
\end{minipage}
\hfill
\begin{minipage}[b]{7.8cm} 
Following \cite{Gattringer:2006ci,Bruckmann:2006kx}
we \emph{twist} the gauge field configuration with a center element as follows: 
all temporal link variables $U_0(\tau,\mbx)$ 
at  a \emph{fixed time} $\tau$ are multiplied with an 
element $z$ in the center $\cZ$ of the
gauge group. The twisted configuration is denoted by
$\{^zU\}$. The Wilson loops $\cW_{\cC}$ for
all \emph{contractable} loops $\cC$ are invariant under
this twisting whereas the Polyakov loops $\cP_\mbx$ pick up the 
center element,
\end{minipage}
\eqnl{
\cW_{\cC}(^zU)=\cW_{\cC}(U)\mtxt{and}
\cP_\mbx(^zU)=z\cP_\mbx (U).}{change}
The Dirac-eigenvalues for the twisted configuration are
denoted by $\lamz_p$. The remarkable and simple 
identity in  \cite{Gattringer:2006ci,Bruckmann:2006kx} relates 
the traced Polyakov loop $L$ to a particular spectral sum,
\eqnl{L=
\frac{1}{\kappa}\sum_{k=1}^{\vert\cZ\vert} \bar z_k\sum_{p=1}^{{\rm dim}(\cD)} \left(\lamzk_p\right)^{N_t},\qquad
 \kappa=(-1)^{\Nt}2^{[d/2]-1} V\vert\cZ\vert.}{gattr1}
The first sum extends over the elements $z_1,z_2,\dots$ in
the center $\cZ$ containing the group identity $e$ 
for which $^{e}\lam_p=\lam_p$. 
The second sum contains the $\Nt$'th power of  all eigenvalues of the
Dirac operator $\dizk$ with twisted gauge fields $\{^{z_k}U\}$. 
It is just the trace or $(\dizk)^{\Nt}$, such that
\eqnl{
L=
\frac{1}{\kappa}\sum_k \bar z_k \,\tr\left(\dizk\right)^{\Nt}
\equiv \Sigma.}{gattr3}
We stress that the formula \refs{gattr3} holds whenever the gauge group
admits a non-trivial center. In \cite{Gattringer:2006ci} it was proved
for $SU(\Nc)$ with center $\Z(\Nc)$ and $\kappa =\frac{1}{2}(-)^{\Nt}\hbox{dim}\cD$. 
In \cite{Bruckmann:2006kx}  the Dirac operator for staggered fermions 
and gauge group $SU(3)$ was investigated and a formula similar to \refs{gattr3} 
was derived. 
Note that \refs{gattr3} is not applicable to the gauge groups $G_2,F_4$ and $E_8$
with trivial centers. 

For completeness we sketch the proof given in \cite{Gattringer:2006ci},
slightly generalized to incorporate all gauge groups with
non-trivial centers. The Wilson-Dirac operator contains hopping
terms between nearest neighbors on the lattice.
A hop from site $x$ to its neighboring site $x\pm e_\mu$ is accompanied 
by the factor $-\frac{1}{2}(1\mp\gamma^\mu)U_\mu(x)$ and
staying at $x$ is accompanied by the factor $m+d$.
Taking the $\ell'$th power of $\cD$,  the single hops combine to chains
of $\ell$ or less hops on the lattice. In particular the trace
$\tr\, \cD^\ell$ is described by loops with \emph{at most} $\ell$ hops.
Each loop $\cC$ contributes a term proportional to the 
Wilson loop  $\cW_\cC$.

On an asymmetric lattice with $\Nt<\Ns$ all loops with length $<\Nt$ are \emph{contractable}
and since the corresponding Wilson loops $\cW_\cC$ do not change under twisting one
concludes
\eqnl{
 \tr\, \diz^\ell=\tr\, \cD^\ell\mtxt{for}\ell<\Nt.}{id1}
For any matrix group with non-trivial $\cZ$ the center
elements sum to zero, $\sum z_k=0$, such that
\eqnl{
\sum_k \bar z_k \tr\big(\dizk^\ell\big)=\tr\big(\cD^\ell\big)\sum_k \bar z_k
=0\mtxt{for}\ell<\Nt.}{id3}
For $\ell=\Nt$ only the Polyakov loops winding once around the periodic time direction are not contractable.
Under a twist by $\{U\}\to \{^zU\}$ they are multiplied by 
$z$, see \refs{change}. With $\sum_k \bar z_k z_k=\vert\cZ\vert$ we end up with
the result \refs{gattr3} which generalizes Gattringer formula to arbitrary 
gauge groups with non-trivial center.
What happens for $\ell>\Nt$ in \refs{id3} will be discussed below.

In \cite{Bruckmann:2006kx} the average shift of the eigenvalues
when one twists the configurations has been calculated. It was observed
that above $T_c$ the shift is greater than below $T_c$ and that the
eigenvalues in the infrared are more shifted than those in the ultraviolet.
But the low lying eigenvalues are relatively suppressed
in the sum \refs{gattr1} such that the main contribution comes from large eigenvalues. 
Indeed, if one considers the \emph{partial sums}
\eqnl{
\Sigma_n=
\frac{1}{\kappa}\sum_k \bar z_k\sum_{p=1}^{n}\lamzk^{N_t}_p,
\quad n\leq \hbox{dim}(\cD),}{gattr5}
where the eigenvalues are ordered according to their absolute values,
then on a $4^3\times 3$-lattice $70\%$ of all eigenvalues 
must be included in \refs{gattr5} to obtain a reasonable approximation to the traced Polyakov 
loop \cite{Bruckmann:2006kx}. Actually, if one includes fewer eigenvalues
then the partial sums have a phase shift of $\pi$ relative to the traced Polyakov loop. 
For large $\Nt$ the contribution from the ultraviolet part of the spectrum
dominates the sum  \refs{gattr1}. Thus it is difficult to see how
the nice lattice result \refs{gattr3} could be of any relevance for 
continuum physics.

The paper is organized as follows:
In the next section we introduce flat connections
with zero curvature but non-trivial Polyakov loops. 
The corresponding eigenvalues of the Wilson-Dirac operator are
determined and spectral sums with support
in the infrared of the spectrum are defined
and computed. The results are useful since they are in 
qualitative agreement with the corresponding results of
Monte-Carlo simulations. In section 3 we recall the construction 
of the real order parameter $L^{\rm rot}$ related to the Polyakov loop
\cite{Wozar:2006fi}. Its Monte-Carlo averages are compared with the 
averages of the partial sums \refs{gattr5}. Our results for
Wilson-Dirac fermions are in qualitative agreement with the corresponding
results for staggered fermions in \cite{Bruckmann:2006kx}.
In section 4 we discuss spectral sums for inverse powers
of the eigenvalues. Their Monte-Carlo averages are proportional
to $\langle L^{\rm rot}\rangle$ such that they are useful
order parameters for the center symmetry. We show
that these order parameters are supported by
the eigenvalues from the infrared end of the spectrum.
Section 5 contains similar results for exponential spectral
sums. Again we find a linear or quadratic relation between
their Monte-Carlo averages and
$\langle L^{\rm rot}\rangle$. It suffices to include only a small
number of infrared eigenvalues in these sums to obtain
efficient order parameters. We hope that the simple relations 
between the infrared-supported spectral sums considered
here and the expectation value $\langle L^{\rm rot}\rangle$ 
are of use in the continuum limit.

\section{Flat connections}\label{flat}
We checked our numerical algorithms against the analytical
results for curvature-free gauge field configurations with non-trivial 
Polyakov loop. For these simple configurations the spatial link variables
are trivial and the temporal link variables are space-independent,
\eqnl{
U_i(x)=\id\mtxt{and}
U_0(x)=U_0(\tau),\quad x=(\tau,\mbx).}{fc1}
The Wilson loops $\cW_\cC$ of all contractable $\cC$ are trivial
which shows that these configurations are curvature-free. We call them
\emph{flat connections}. With the gauge transformation
\eqnl{
\Omega(\tau)=\cP^{-1}_\tau,\quad \cP_\tau=U_0(\tau-1)U_0(\tau-2)\cdots U_0(2)U_0(1) }{fc3}
all link-variables of a flat connection are transformed into the group-identity. 
But the transformed fermion fields are not periodic in time anymore,
\eqnl{
\psi(\tau+\Nt,\mbx)=\cP^{-1}\psi(\tau,\mbx),\mtxt{where} \cP=\cP_{\Nt+1}}{fc5}
is just the constant Polyakov loop. Since the transformed Dirac operator
is the free operator, its eigenfunctions are plane waves,
\eqnl{
\psi(x)=e^{ipx}\psi_0.}{fc7}
These are eigenmodes of  the free Wilson-Dirac operator with eigenvalues
$\{\lam_p\}=\{\lam_p^\pm\}$, where
\eqnl{
\lam^\pm_p=m\pm i\vert \ring{p}\vert+\frac{r \hat{p}^2}{2},\mtxt{with}
\hat p_\mu=2\sin \frac{p_\mu}{2},\quad \ring{p}_\mu=\sin p_\mu.}{fc9}
They are periodic in the space directions provided the spatial momenta
are from
\eqnl{
p_i\in \frac{2\pi}{N_s}n_i\mtxt{with} n_i\in \Z_{\Ns}.}{fc11}
Denoting the eigenvalues of the Polyakov loop by
$e^{2\pi i\varphi_1},\dots ,e^{2\pi i\varphi_{N_c}}$, the periodicity
conditions \refs{fc5} imply
\eqnl{
p_0=\frac{2\pi}{N_t}(n_0-\varphi_j),\quad n_0\in \Z_{\Nt},\quad j=1,\dots,\Nc.}{fc13}
Thus the eigenvalues of the Wilson-Dirac operator with a flat connection
are given in \refs{fc9}, with quantized momenta \refs{fc11} 
and \refs{fc13}. For each momentum $p_\mu$ there exist $2^{[d/2]-1}$ eigenvalues 
$\lam_p^+$ and  $2^{[d/2]-1}$ complex conjugated eigenvalues $\lam_p^-$. 

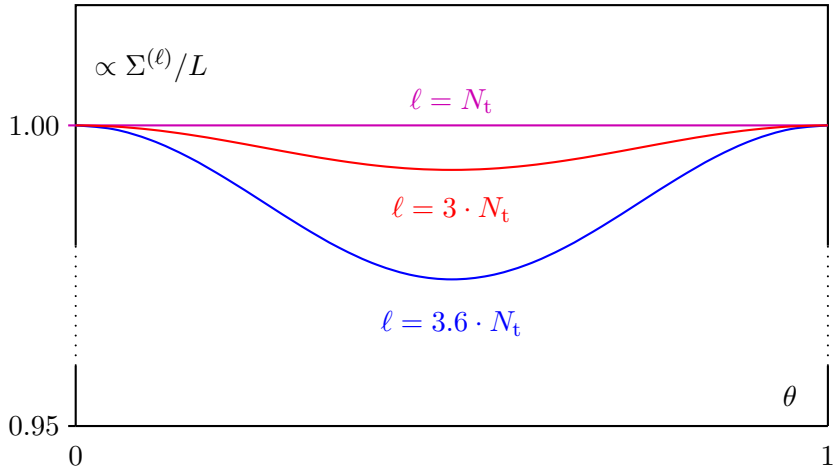
\begin{figure}[ht]
\psset{xunit=1mm,yunit=8mm,arrowsize=1.4mm}
\begin{pspicture}(-5,94.5)(105,102.5)
\psline(0,98)(0,102)(100,102)(100,98)
\psline[linestyle=dotted](0,98)(0,96)
\psline[linestyle=dotted](100,98)(100,96)
\psline(0,95)(100,95)
\small
\rput(95,95.5){$\theta$}
\psline(100,95)(100,96)\rput(100,94.5){$1$}
\psline(0,95)(0,96)\rput(0,94.5){$0$}
\psline[linecolor=wpurple](-1,100)(100,100)
\rput(-5.5,100){$1.00$}
\psline(-1,95)(0,95)\rput(-5.5,95){$0.95$}
\rput[l](2.5,101){$\propto \Sigma^{(\ell)}/L$}
\rput(50,100.4){\wpurple $\ell=\Nt$}
\rput(50,98.6){\red $\ell=3\cdot\Nt$}
\rput(50,96.7){\blue $\ell=3.6\cdot \Nt$}
\pscurve[linecolor=blue]
(0,100)(5,99.937)(10,99.755)(15,99.472)(20,99.115)(25,98.720)(30,98.324)(35,97.967)
(40,97.684)(45,97.502)(50,97.439)(55,97.502)(60,97.684)(65,97.967)(70,98.324)
(75,98.720)(80,99.115)(85,99.472)(90,99.755)(95,99.937)(100,100)
\pscurve[linecolor=red]
(0,100)(5,99.982)(10,99.929)(15,99.847)(20,99.744)(25,99.630)(30,99.516)(35,99.413)
(40,99.331)(45,99.278)(50,99.260)(55,99.278)(60,99.331)(65,99.413)(70,99.516)
(75,99.630)(80,99.744)(85,99.847)(90,99.929)(95,99.982)(100,100)
\end{pspicture}
\caption{\label{fig:flatconnection1} Spectral sums
$\Sigma^{(\ell)}$ divided by the traced Polyakov loop as functions of $\theta$
for different values of $\ell$.}
\end{figure}

Next we twist the flat connections with a center-element,
for $SU(N_c)$ with
\eqnl{
z_k=e^{2\pi i k/N_c}\id,\quad 1\leq k\leq N_c.}{fc15}
The spatial components of the momenta are still given by \refs{fc11}, but their
temporal component is shifted by an amount proportional to $k$,
\eqnl{
p_0(z_k)\in\left\{\frac{2\pi}{\Nt}\left(n_0-\varphi_j -k/\Nc\right)\right\},\quad 
1\leq j,k\leq \Nc.}{fc17}
In the following we consider flat $SU(3)$-connections with Polyakov 
loops
\eqnl{
\cP(\theta)=\pmatrix{e^{2\pi i\theta}&0&0\cr
0&1&0\cr 0&0&e^{-2\pi i\theta}}\Longrightarrow
L=1+2\cos(2\pi\theta).}{fc19}
For these fields the temporal component of the momentum takes values from
\eqnl{
p_0(z_k)\in\left\{\frac{2\pi}{\Nt}\left(n_0-j \theta -k/3\right)\right\},\quad j\in\{-1,0,1\},\quad
k\in\{0,1,2\}.}{fc21}
We have calculated the spectral sums
\eqnl{
\Sigma^{(\ell)}=
\frac{1}{\kappa}\sum_k \bar z_k\sum_{p=1}^{{\rm dim} \cD}
\left(\lamzk_p\right)^{\ell}
=\frac{1}{\kappa}\sum_k \bar z_k \,\tr\left( \dizk\right)^\ell}{fc23}
for vanishing mass. For flat connections
the sums with powers $\ell$ between $\Nt$ and $2\Nt$ 
are strictly proportional to the traced Polyakov loop, 
$\Sigma^{(\ell)}=C_\ell L(\theta)$. Gattringers result implies $C_{N_t}=1$. The next 
two coefficients are related to the number of loops of
length $\Nt+1$ and $\Nt+2$ winding once around the periodic 
time direction. One finds
\eqnl{
C_{\Nt+1}=d(\Nt+1)\mtxt{and}
C_{\Nt+2}=\frac{d^2}{2}(\Nt+2)(\Nt+1)+
\frac{d-1}{4}\big(\Nt(\Nt+1)-2\big).}{fc23a}
More generally, the relation $\sum \bar z_k z_k^\ell=0$ for $\ell \notin 3\Z+1$
implies that the spectral sums \refs{fc23} are linear combinations of 
the traces $\tr \,\cP^{3n+1}(\theta)$ for sufficiently small values of $\vert 3n+1\vert $,
\eqnl{
\Sigma^{(\ell)}=\sum_{n:\,\vert 3n+1\vert \Nt\leq \ell} C^{(n)}_\ell\, \tr\, \cP^{3n+1}(\theta).}{fc23b}
In Fig. \ref{fig:flatconnection1} we depicted the sums $\Sigma^{(\ell)}$ on
a $4\times 12^3$ lattice, divided by
the traced Polyakov loop and normalized to one for $\theta=0$
for the flat connections and the powers $\ell=\Nt,\, 3\Nt$ and $3.6\Nt$.
Note that the power $\ell$ in \refs{fc23} need not be an integer.

We have argued that the sum $\Sigma^{(\ell)}$ must be a linear combination of 
$\tr\,\cP$ and $\tr\,\cP^{-2}$ for $\ell$ between $2\Nt$ and $4\Nt$.
Actually, up to $\ell\approx 3\Nt$ the sum is well approximated by a 
multiple of $\tr\, \cP$. This is explained by the fact that for a given $\ell$ there are 
much more fat loops winding once around the periodic time direction 
and contributing with $\tr\,\cP$ than there are thin long loops winding many times 
around and contributing with $\tr\, \cP^{-2},\tr\,\cP^4,\,\tr\,\cP^{-5},\dots$. We shall 
see that similar results apply to the expectation values of  $\Sigma^{(\ell)}$ 
in Monte-Carlo generated ensembles of gauge fields.

Since the eigenvalues in the infrared are mostly affected by the twisting \cite{Bruckmann:2006kx}
we could as well choose a spectral sum for which the 
ultraviolet end of the spectrum is suppressed. Since $\Sigma^{(\ell)}$ 
with $\ell\leq 3\Nt$ is almost proportional to the traced Polyakov loop there exist
many such spectral sums. They define order parameters for
the center symmetry and may possess a well-defined continuum limit. 
For example,  the exponential sums
\eqnl{
\cE^{(\ell)}=
\frac{1}{\kappa}\sum_k \bar z_k\sum_{p=1}^{{\rm dim} \cD}
e^{-\ell\cdot \lam_p\left(^{z_k} U\right)},}{fc25}
are all proportional to the traced Polyakov loop for a factor $\ell$
in the exponent between $0.1$ and $2$. Below we displayed
exponential sums for the flat connections on
a $4\times 12^3$-lattice and various $\ell$ between $0.1$ and
$2$. Again we divided by the traced Polyakov loop $L(\theta)$
and normalized the result to unity for $\theta=0$.
\begin{figure}[ht]
\psset{xunit=1mm,yunit=4.5mm}
\begin{pspicture}(-5,88)(105,104)
\psframe(0,90)(100,104)
\small
\rput(100,89){$1$}
\rput(0,89){$0$}
\rput(96,90.8){$\theta$}
\rput(-5,100){$1.0$}
\psline(-1,90)(0,90)\rput(-5,90){$0.9$}
\psline[linecolor=wpurple](-1,100)(100,100)
\rput[l](2.2,102.8){$\propto \cE^{(\ell)}/L$}
\rput(50,99){\wpurple $\ell=1$}
\rput(50,100.8){\blue $\ell=0.1$}
\rput(50,92){\red $\ell=2$}
\pscurve[linecolor=wpurple]
(0,100)(5,99.993)(10,99.974)(15,99.944)(20,99.906)(25,99.864)(30,99.822)(35,99.784)
(40,99.754)(45,99.734)(50,99.728)(55,99.734)(60,99.754)(65,99.784)(70,99.822)
(75,99.864)(80,99.906)(85,99.944)(90,99.974)(95,99.993)(100,100)
\psline[linecolor=blue](0,100)(100,100)
\pscurve[linecolor=red]
(0,100)(5,99.833)(10,99.349)(15,98.595)(20,97.645)(25,96.591)(30,95.538)(35,94.588)
(40,93.834)(45,93.350)(50,93.183)(55,93.350)(60,93.834)(65,94.588)(70,95.538)
(75,96.591)(80,97.645)(85,98.595)(90,99.349)(95,99.833)(100,100)
\end{pspicture}
\caption{\label{fig:flatconnection3} Spectral sums
$\cE^{(\ell)}$ divided by the traced Polyakov loop as functions of $\theta$
for different values of $\ell$.}
\end{figure}
\\
Later when we use  Monte-Carlo generated configurations 
to calculate the expectation values of $L$ and $\cE^{(\ell)}$ we shall 
choose $\ell=1$. For this choice the mean exponential sum will be 
proportional to the mean $L$. Later we shall argue why this is the case.
\section{Distribution of Dirac eigenvalues for SU(3)}\label{distribution}
We have undertaken extended numerical studies of the
eigenvalue distributions and various spectral sums
for the Wilson-Dirac operator in $SU(3)$ lattice gauge theory. 
First we summarize our results on the partial traces
\eqnl{
\Sigma_n^{(\ell)}=
\frac{1}{\kappa}\sum_k \bar z_k\sum_{p=1}^{n}\lamzk^{\ell}_p\,,
\qquad n\leq \hbox{dim}(\cD),\quad \vert\lam_p\vert\leq \vert\lam_{p+1}\vert.}{distr1}
For $n=\hbox{dim}(\cD)$ one sums over all eigenvalues of the Dirac-operator
and obtains the traces $\Sigma^{(\ell)}$ considered in
\refs{fc23}. For $\ell=\Nt$ one finds the partial sums $\Sigma_n$ in
\refs{gattr5}. These have been 
extensively studied for staggered fermions in \cite{Bruckmann:2006kx}.
According to the result \refs{gattr1} the object  $\Sigma_{{\rm dim}\cD}$ is just the
traced Polyakov loop. 

\begin{figure}[ht]
\includegraphics{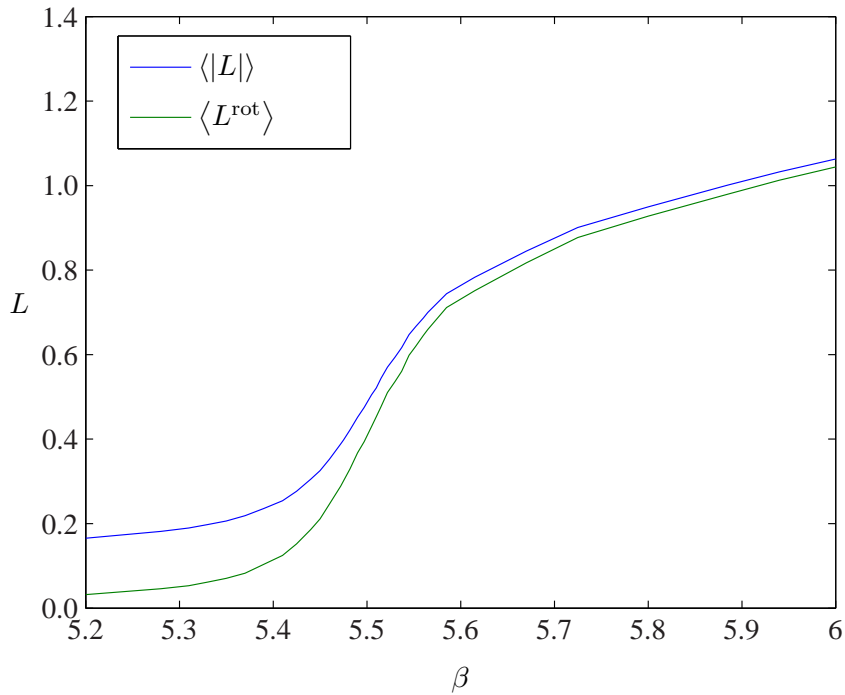}
\caption{\label{fig:loopsOverBeta} Dependence of the mean modulus of
$L$ and the center-transformed and rotated $L$ (see text) on the 
Wilson coupling $\beta$ on a $4^3\times 3$ lattice. The critical coupling is $\beta_{\rm crit}=5.49$}
\end{figure}

\begin{figure}[ht]
\includegraphics{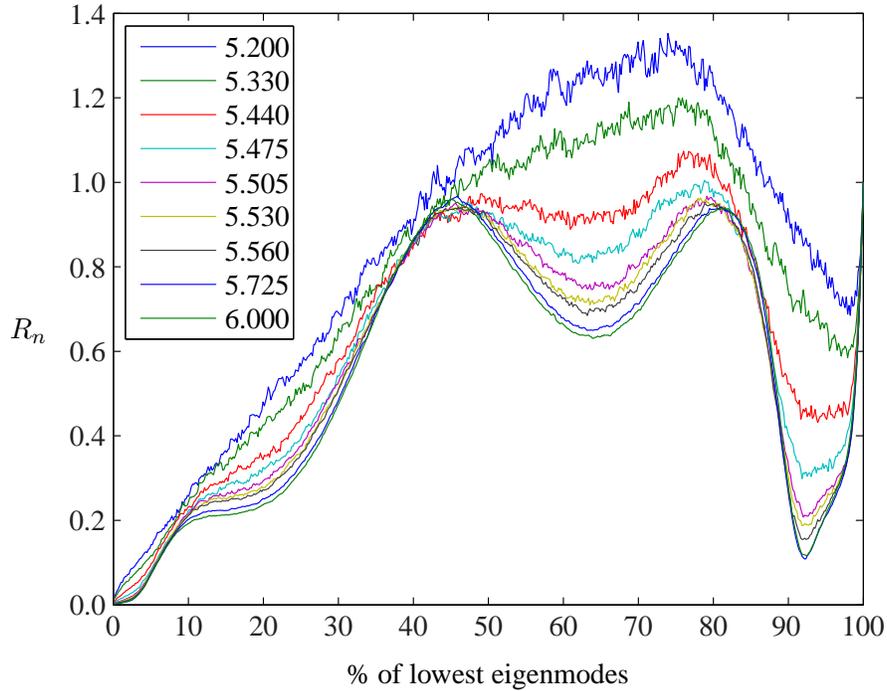}
\caption{\label{fig:curvesAbs43} Modulus of the eigenvalue sums starting from
the lowest eigenmodes on a $4^3\!\times\!3$-lattice near the phase transition.
The distinct graphs are labelled with the Wilson coupling $\beta$.}
\end{figure}

\begin{figure}[ht]
\includegraphics{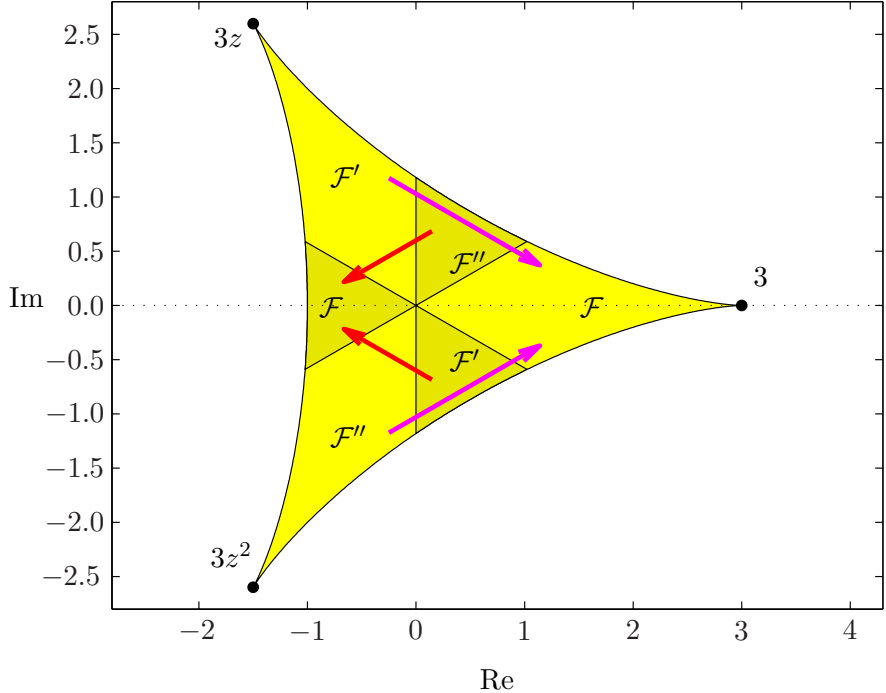}
\caption{\label{fig:domainMapping} Fundamental domain $\cF$
of $L$ obtained by identifying $\Z(3)$
copies according to the depicted arrows.}
\end{figure}

We did simulations on lattices with sizes up to $8^3\times 4$.
Here we report on the results obtained on a $4^3\times 3$ lattice
with critical coupling $\beta_{\rm crit}\approx 5.49$,
determined with the histogram method based on $40\,000$ configurations.
The dependence of the two order parameters $\vert L\vert$ and $L^{\rm rot}$ 
(see below) on the Wilson coupling $\beta$ has been calculated for
$35$ different $\beta$ and is depicted in Fig. \ref{fig:loopsOverBeta}.
For each $\beta$ between $4\,000$ and $20\,000$  independent  configuration 
have been generated  and analyzed. For our relatively small lattices the two order 
parameters change gradually
from the symmetric confined to the broken deconfined phase.
Table \ref{table:loopsOverBeta} contains the order parameters for $11$ Wilson
couplings.
\begin{table}[ht]
\begin{tabular}{|c|ccccc|cccccccc|}\hline
$\beta$&5.200& 5.330 & 5.440& 5.475&&&  5.505& 5.530& 5.560& 5.615 
&  5.725& 5.885&6.000\\ \hline
 $\langle\vert L\vert\rangle$&0.1654& 0.1975 & 0.3050&0.3980&&& 0.5049 & 0.5939
 &0.6865 &0.7832  & 0.9007 &1.0013&1.0631  \\
$\langle L^{\rm rot}\rangle$
&0.0318& 0.0615 & 0.1859 &0.3013&&& 0.4296 &0.5363  &0.6452 & 0.7513
&0.8770 &0.9797& 1.0444
\\ \hline
\end{tabular}  
\caption{\label{table:loopsOverBeta} Dependence of the order parameters
$\vert L\vert $ and $L^{\rm rot}$ on the 
Wilson coupling $\beta$.}
\end{table}
For every independent configuration we calculated the dim$(\cD)=2304$ eigenvalues 
of the Wilson-Dirac operator.
Then we averaged the absolute values of the partial traces $\Sigma_n$ for every $\beta$ 
in table \ref{table:loopsOverBeta}. In Fig. \ref{fig:curvesAbs43} the ratios 
\eqnl{
R_n=
\frac{\langle\vert \Sigma_n\vert\rangle}{\langle \vert L\vert\rangle}}{mn1}
are plotted for these $\beta$ as function of the percentage of 
eigenvalues considered in the partial traces.
\\
To retain information on the phase of the partial traces and Polyakov loop
we used the invariant order parameter constructed in \cite{Wozar:2006fi}.
Recall that the domain for the traced Polyakov loop is the triangle shown 
in Fig. \ref{fig:domainMapping}. The three elements in the center of $SU(3)$ 
correspond to the corners of the triangle.
We divide the domain into the six distinct
parts in Fig.~\ref{fig:domainMapping}. The light-shaded
region represents the preferred location of the traced Polyakov loop $L$ in the
deconfined (ferromagnetic) phase, whereas the dark-shaded region corresponds
to the hypothetical anti-center ferromagnetic phase \cite{Wipf:2006wj}. 
In the deconfined phase $L$ points in the direction of a center 
element whereas it points in the opposite direction in the anti-center phase.
To eliminate the superfluous center-symmetry
we identify the regions as indicated by the
arrows in Fig.~\ref{fig:domainMapping}. This way we end up with a
\emph{fundamental domain} $\cF$ for the center-symmetry
along the real axis. Every $L$ is mapped into
$\cF$ by a center transformation. To finally obtain a real
observable we rotate the transformed $L$ inside $\cF$ onto the real axis. The
result is the variable $L^{\rm rot}$ whose sign
clearly distinguishes between the different phases.
$L^{\rm rot}$ is negative in the anti-center phase, positive in the
deconfined phase and zero in the confined symmetric phase.
The object $L^{\rm rot}$ is a useful order parameter for
the confinement-deconfinement phase transition in gluodynamics
\cite{Wozar:2006fi}. 

We performed the same construction with
the partial sums $\Sigma_{n}$ and calculated the ratios for the corresponding 
Monte-Carlo averages 
\eqnl{
R^{\rm rot}_n=\frac{\langle \Sigma^{\rm rot}_n\rangle}{\langle L^{\rm rot}\rangle}
}{mn3}
for every $\beta$ in table  \ref{table:loopsOverBeta} as a function of the percentage of 
eigenvalues considered in $\Sigma_n$.
\begin{figure}[ht]
\includegraphics{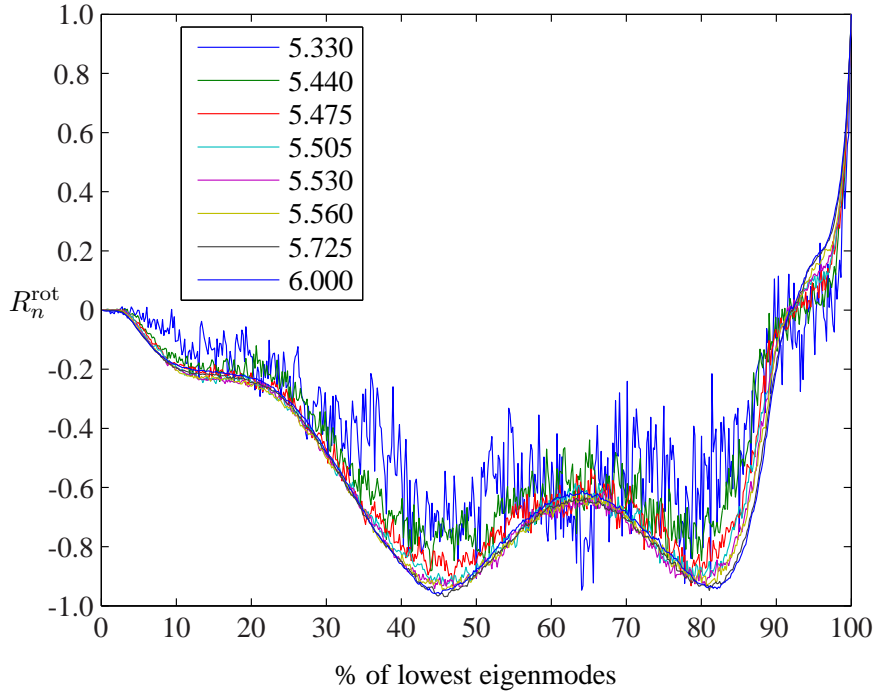}
\caption{\label{fig:curvesRot43} Eigenvalue sums rotated to the fundamental
domain starting from the lowest eigenmodes on a $4^3\!\times\!3$-lattice near 
the phase transition. The distinct graphs are labelled with the Wilson coupling.}
\end{figure}
In Fig. \ref{fig:curvesAbs43} and \ref{fig:curvesRot43} we observe a universal
behavior in the deconfined phase with modulus of the traced Polyakov loop larger than
approximately $0.4$. If we include less than $90\% $ of the eigenvalues, then
the partial sums $\Sigma_n$ have a phase shift of $\pi$ in comparison with 
$\Sigma=\Sigma_{\rm dim \cD}$.
The last dip in Fig \ref{fig:curvesAbs43} is due to this phase shift and 
indicates the  transition through zero that occurs  when $\Sigma_n$ changes sign.
The same shift and dip has been reported for staggered fermions 
on a $6^3\times 4$ lattice in \cite{Bruckmann:2006kx}.
For staggered fermions $\Sigma_n$ and $\Sigma$ are
in phase for $n\geq 0.65\cdot \hbox{dim}(\cD)$.  For Wilson-Dirac fermions 
this happens only for $n\geq 0.9\cdot\hbox{dim}(\cD)$.
\paragraph{Finite spatial size scaling of partial sums:}\label{Finite}
We fixed the coupling at $\beta=5.5$ and simulated in the deconfined
phase on $\Ns^3\times 2$-lattices with varying spatial sizes $\Ns\in\{3,4,5\}$.
For this coupling the systems are deep in the broken phase and we 
can study finite size effects on the spectral sums.
\begin{figure}[ht]
\includegraphics[scale=0.95]{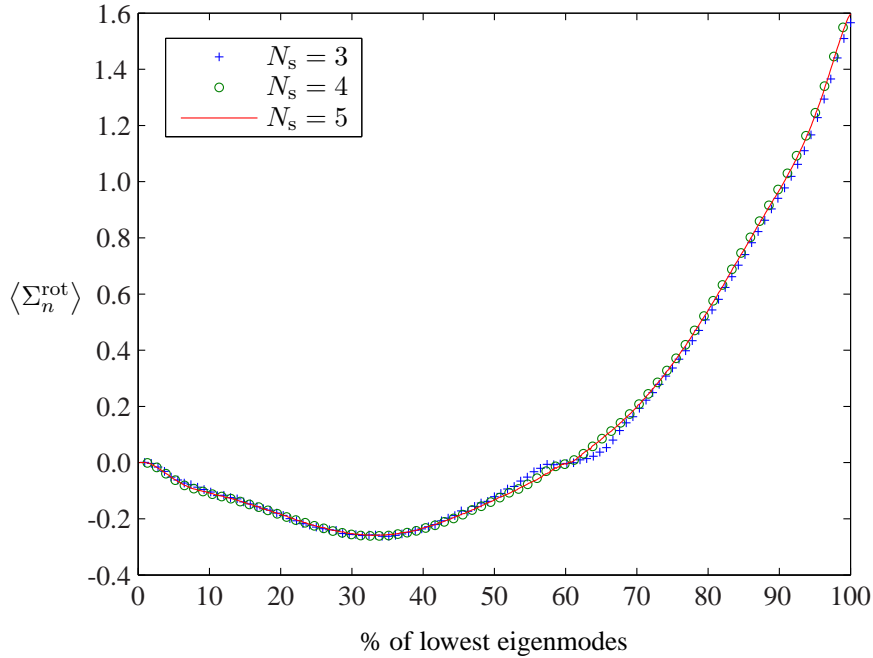}
\caption{\label{fig:scalingCurves}  Rotated eigenvalue sums
starting from the lowest $\lam_p$ on a $\Ns^3\!\times\!2$-lattice 
in the broken phase.}
\end{figure}
The results for $\Sigma_n^{\rm rot}$ are depicted in Fig. \ref{fig:scalingCurves}.
We observe that  to a high precision $\Sigma_n^{\rm rot}$ is approximately 
independent of the spatial volume. The curves for $\Ns=4$ and $5$ are not 
distinguishable in the plot and as expected $\sum_k \bar z_k \tr( \dizk^{\Nt})$
scales with the spatial volume of the system. An increase of $\Ns$ affects 
the spectra for the untwisted and twisted configurations alike --
they only become denser with increasing spatial volume.
On the other hand, comparing Fig. \ref{fig:curvesRot43} and Fig. \ref{fig:scalingCurves}, 
it is evident that the graph of $\Sigma_n^{\rm rot}$ depends very much on the temporal 
extent of the lattice. 
\paragraph{Partial traces $\Sigma_n^{(\ell)}$:}
The truncated eigenvalue sums \refs{distr1} with
different powers $\ell$ of the eigenvalues show an universal behavior 
that is nearly independent of the lattice size. The main reason
for this universality and in particular the sign of $\Sigma_n^{(\ell)}$ 
is found in the response of the low-lying eigenvalues to twisting
the gauge field. It has been observed that for non-periodic
boundary conditions (which are gauge-equivalent to twisting the gauge
field) the low lying eigenvalues are on the average further away from the 
origin as compared to periodic boundary conditions (or untwisted gauge
fields) \cite{Chandrasekharan:1995gt,Chandrasekharan:1995nf,Stephanov:1996he,Gattringer:2002dv}.
This statement is very clear for massless staggered fermions with
eigenvalues on the imaginary axis. For example, the partial traces
\eqnl{
\Sigma_n^{(1)}
\propto
\sum_{p=1}^n\lam_p+\bar z \sum_{p=1}^n \lamz_p+z \sum_{p=1}^n  {^{\bar z}\lam}_p,\qquad
\vert \lam_p\vert\leq \vert \lam_{p+1}\vert.}{mn5}
with $n\ll \hbox{dim}\,\cD$ and the traced Polyakov loop
have opposite phases. This is explained as follows:
all sums in \refs{mn5} are positive and on the average the last
two sums are equal. With $z+\bar z=-1$ the last two terms 
add up to $-\sum \lamz_p$. Since the low lying eigenvalues
for the twisted field are further away from the origin as
for the untwisted field, the spectral sums \refs{mn5} are negative
for small $n$. 
\section{Traces of propagators}\label{traces}
\begin{figure}[ht]
\includegraphics{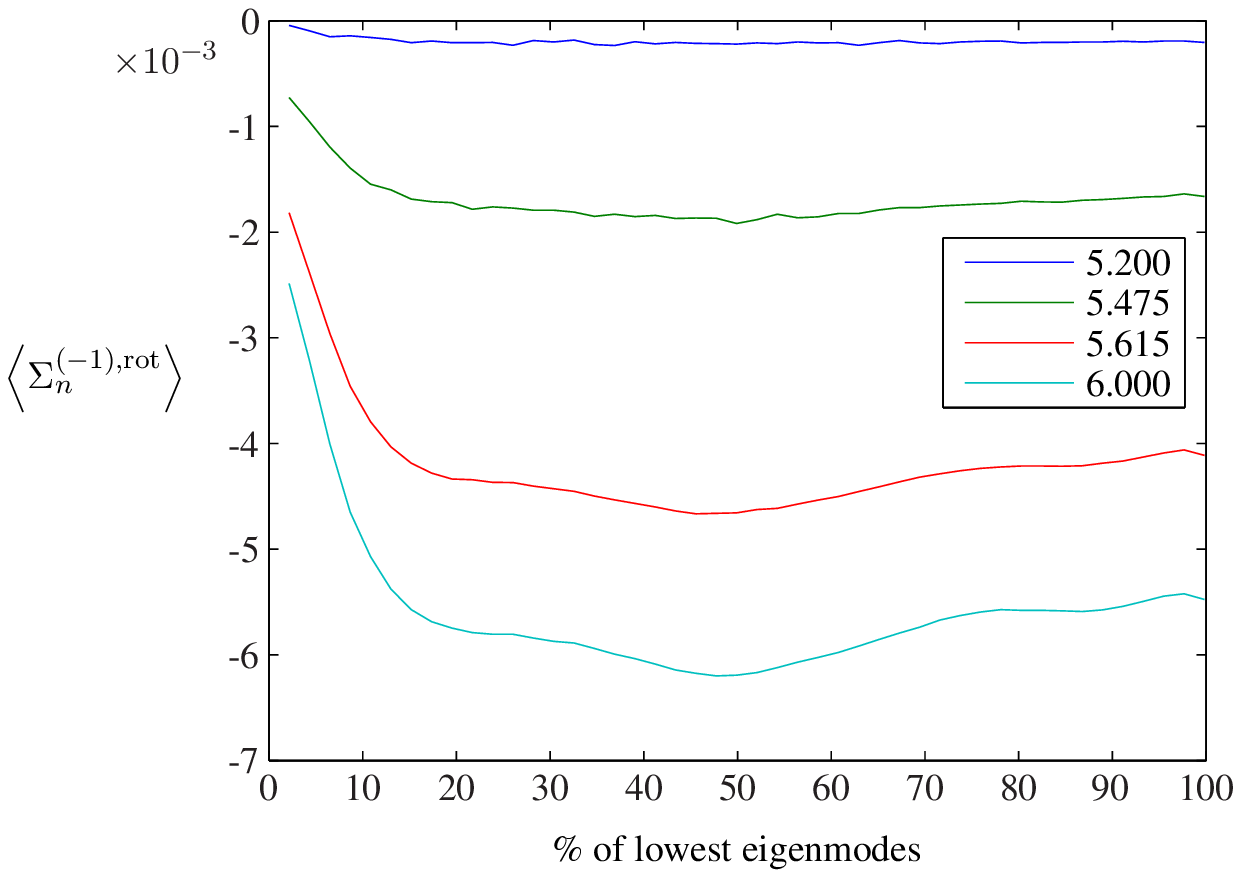}
\includegraphics{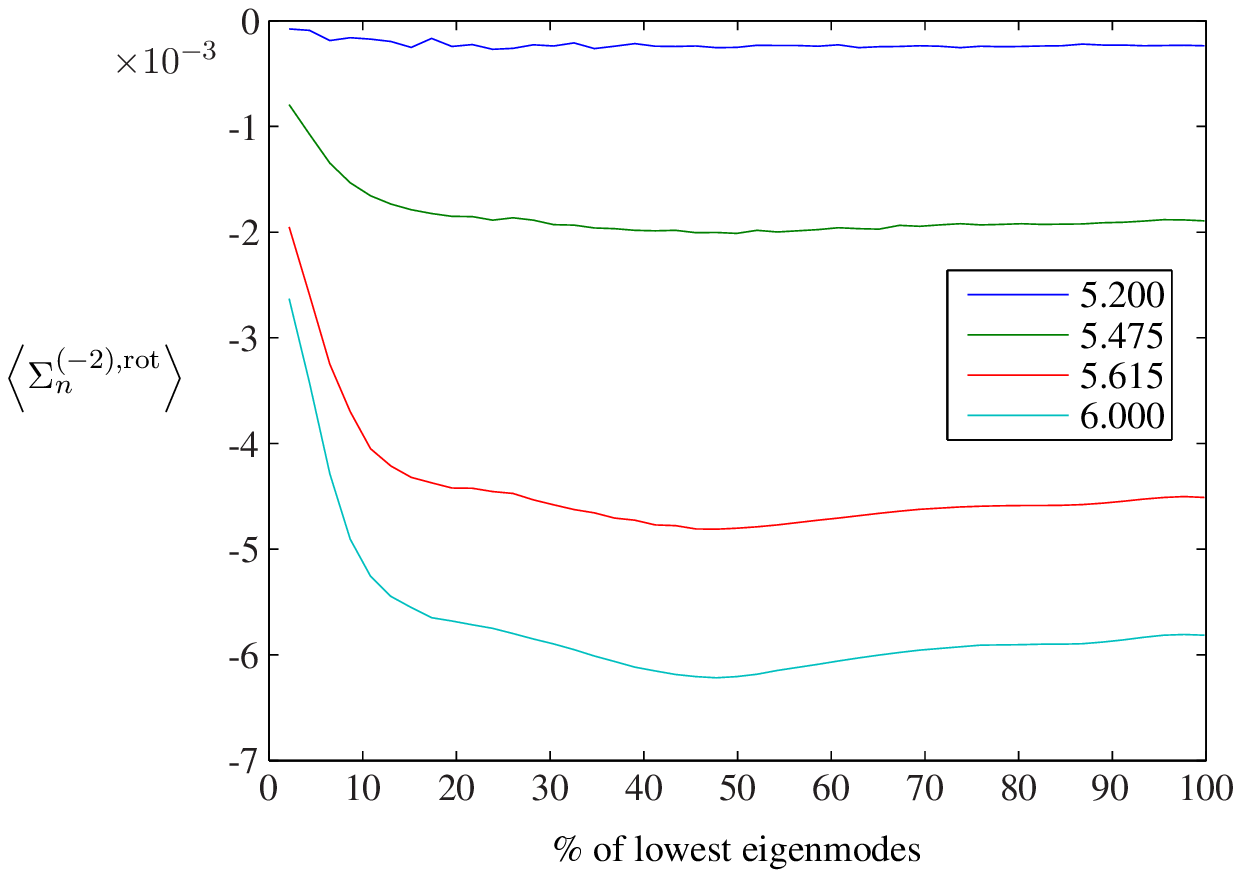}
\caption{\label{fig:lambdaMinusEinsZwei43} 
The expectation values of the partial spectral sums $\Sigma^{(-1)}_n$ and $\Sigma^{(-2)}_n$ rotated to the fundamental domain starting from the lowest eigenvalue on a $4^3\times 3$ lattice. The
graphs are labelled with $\beta$.}
\end{figure}

To suppress the contributions of large eigenvalues we introduce 
spectral sums $\Sigma^{(\ell)}$ with \emph{negative exponents} $\ell$. Similar to
the Polyakov loop these sums serve as order parameters for the center
symmetry. In particular the spectral sums
\eqnl{
\Sigma^{(-1)}=\frac{1}{\kappa}\sum_k \tr \left(\frac{\bar z_k}{\dizk}\right)\mtxt{and}
\Sigma^{(-2)}=\frac{1}{\kappa}\sum_k \tr \left(\frac{\bar z_k}{\dizk^2}\right)\
}{mn7}
are of interest, since they relate to the Green functions of $\cD$
and $\cD^2$, objects which enter the discussion
of the celebrated Banks-Casher relation.  Contrary to the 
ultraviolet-dominated sums with positive $\ell$ 
are the sums with negative $\ell$ dominated by the eigenvalues
in the infrared. The operators $\{\diz,\,z\!\in\!\cZ\}$ have similar spectra 
and we may expect that $\kappa\Sigma^{(-1)}$ has
a well-behaved continuum limit. 
Here we consider the partial  traces 
\eqnl{
\Sigma^{(-1)}_n=
\frac{1}{\kappa}\sum_k \bar z_k\sum_{p=1}^{n}\frac{1}{\lamzk_p},
\mtxt{and}
\Sigma^{(-2)}_n=
\frac{1}{\kappa}\sum_k \bar z_k\sum_{p=1}^{n}\frac{1}{(\lamzk_p)^2},\quad 
\vert\lam_p\vert\leq \vert\lam_{p+1}\vert.}{mn8}
Since the Wilson-Dirac operator with flat connection possesses zero-modes we
added a small mass $m=0.1$ to the denominators in \refs{mn8}. 
\begin{figure}[ht]
\psset{xunit=4mm,yunit=2mm}
\begin{pspicture}(-2,2)(26,-38)
\rput(4,-2){$1000\cdot \Sigma^{(-1)}_n$}
\rput(12,-39.5){$\%$ of lowest eigenvalues}
\small
\rput(1,-36.5){0}\rput(23,-36.5){100}
\psline(12,-35)(12,-34)\rput(12,-36.5){50}
\psline(6.5,-35)(6.5,-34)\rput(6.5,-36.5){25}
\psline(17.5,-35)(17.5,-34)\rput(17.5,-36.5){75}
\rput[r](.5,0){$0$}
\psline(1,-10)(1.5,-10)\rput[r](.5,-10){$-10$}
\psline(1,-20)(1.5,-20)\rput[r](.5,-20){$-20$}
\psline(1,-30)(1.5,-30)\rput[r](.5,-30){$-30$}
\rput(20,-6.3){\wblue $L=.25$}
\rput(20,-16.5){\red $L=.50$}
\rput(20,-23.7){\wpurple $L=.75$}
\rput(20,-28.7){$L=1.0$}
\psframe(1,0)(23,-35)
\pscurve[linecolor=wbl](1,-6.346)(2,-7.016)(3,-7.047)(4,-7.273)(5,-7.329)(6,-7.343)(7,-7.070)
(8,-7.777)(9,-7.525)(10,-7.553)(11,-7.375)(12,-7.319)(13,-7.670)(14,-7.312)
(15,-7.433)(16,-7.396)(17,-7.420)(18,-7.493)(19,-7.280)(20,-7.364)(21,-7.365)
(22,-7.307)(23,-7.330)
\pscurve[linecolor=red](1,-15.667)(2,-17.020)(3,-17.094)(4,-17.297)(5,-17.646)(6,-17.686)(7,-18.916)
(8,-18.135)(9,-18.041)(10,-18.091)(11,-17.748)(12,-17.643)(13,-18.272)(14,-17.633)
(15,-17.866)(16,-17.824)(17,-18.104)(18,-17.849)(19,-17.571)(20,-17.731)(21,-17.731)
(22,-17.620)(23,-17.665)
\pscurve[linecolor=wpurple]
(1,-21.855)(2,-23.919)(3,-24.046)(4,-24.408)(5,-24.850)(6,-24.939)(7,-26.269)
(8,-25.295)(9,-25.314)(10,-25.308)(11,-25.045)(12,-25.454)(13,-25.525)(14,-25.444)
(15,-25.144)(16,-25.062)(17,-25.388)(18,-25.125)(19,-24.790)(20,-25.041)(21,-25.001)
(22,-24.874)(23,-24.906)
\pscurve
(1,-25.511)(2,-28.338)(3,-29.003)(4,-28.949)(5,-29.654)(6,-30.860)(7,-31.143)
(8,-30.048)(9,-30.206)(10,-30.157)(11,-29.954)(12,-30.313)(13,-30.308)(14,-30.349)
(15,-29.974)(16,-29.971)(17,-30.218)(18,-29.950)(19,-29.854)(20,-29.834)(21,-29.736)
(22,-29.677)(23,-29.676)
\end{pspicture}
\caption{\label{fig:inverseDirac}The partial spectral sums $\Sigma^{(-1)}_n$
for the inverse power and flat connections.}
\end{figure}
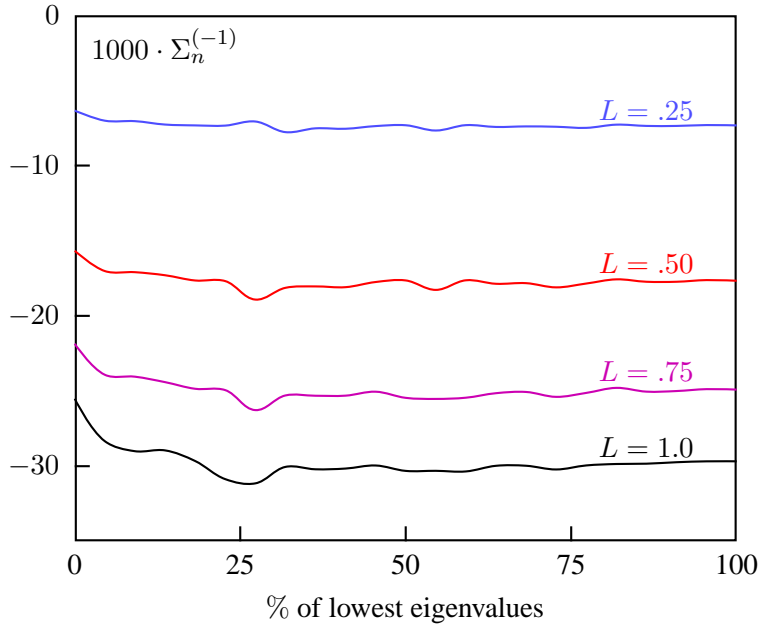
In Fig. \ref{fig:inverseDirac} the partial sums $\Sigma^{(-1)}_n$ on a $4^3\times 3$ lattice
are plotted. 
It is seen that for flat connections the $\Sigma^{(-1)}_n$ for
small $n$ are excellent indicators for  
the traced Polyakov loop. Thus it is tempting to propose $\Sigma^{(-1)}_n,\,\Sigma^{(-2)}_n$ with
$n\ll {\rm dim}\cD$ as order parameters for the center symmetry. To test this
proposal we calculated the partial sums \refs{mn8}, transformed
to the fundamental domain and rotated to the real axis,
for Monte-Carlo  generated configurations on a $4^3\times 3$ lattice for various values of
$\beta$. The results in Fig. \ref{fig:lambdaMinusEinsZwei43}
are qualitatively similar to those for the flat connections.
Taking into account $10\%$ of the eigenvalues in the IR already yields the
asymptotic values $\Sigma^{(-1),{\rm rot}}$ and $\Sigma^{(-2),{\rm rot}}$.

\begin{figure}[ht]
\includegraphics{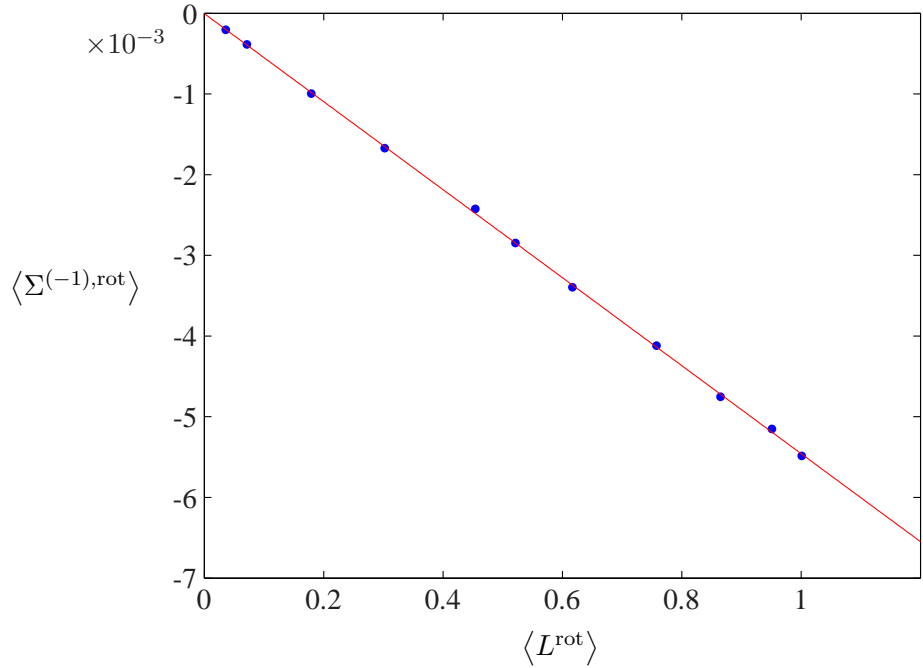}
\caption{\label{fig:propLambdaMinusEins43.eps} 
The expectation values of $\Sigma^{(-1),\rm rot}$
as functions of $\langle L^{\rm rot}\rangle$ on a $4^3\times 3$ lattice.}
\end{figure}

To find an approximate relation between $\Sigma^{(-1)}$ and the traced
Polyakov loop we applied the hopping-parameter expansion. To that end one
expands the inverse of the Wilson-Dirac operator $\cD=(m+d)\id -V$ in powers of $V$,
\eqnl{
\cD^{-1}=\frac{1}{m+d}\sum_k \frac{1}{(m+d)^k}\big[(m+d)\id -\cD\big]^k.
}{mn9}
Inserting this Neumann series into $\Sigma^{(-1)}$ in \refs{mn7} and keeping
the leading term only yields
\eqnl{
\Sigma^{(-1)}= \frac{(-1)^{\Nt}}{(m+d)^{\Nt+1}}\,\Sigma^{(1)}+\dots
\stackrel{\refs{gattr3}}{\approx}\frac{(-1)^{\Nt}}{(m+d)^{\Nt+1}}L.}
{mn11}
To check whether the expectation   
values of $\Sigma^{(-1),\rm rot}$ and $L^{\rm rot}$ are indeed proportional to 
each other we have calculated these values for Monte-Carlo
ensembles corresponding to the $11$ Wilson couplings in table \ref{table:loopsOverBeta}.
The results in Fig. \ref{fig:propLambdaMinusEins43.eps} clearly
demonstrate that there is such a linear relation.
\\
A linear fit yields 
\eqnl{
\langle\Sigma^{(-1),\rm rot}\rangle=-0.00545\cdot \langle L^\mathrm{rot}\rangle-4.379\cdot 10^{-6}
\quad ({\rm rmse} = 2.978\cdot 10^{-5}).}{mn13}
For massless fermions on a $4^3\times 3$ lattice the crude approximation \refs{mn1}
leads to a slope $-0.003906$.  This is not far off the slope $-0.00545$ extracted 
from the Monte-Carlo data. 

\begin{figure}[ht]
\includegraphics{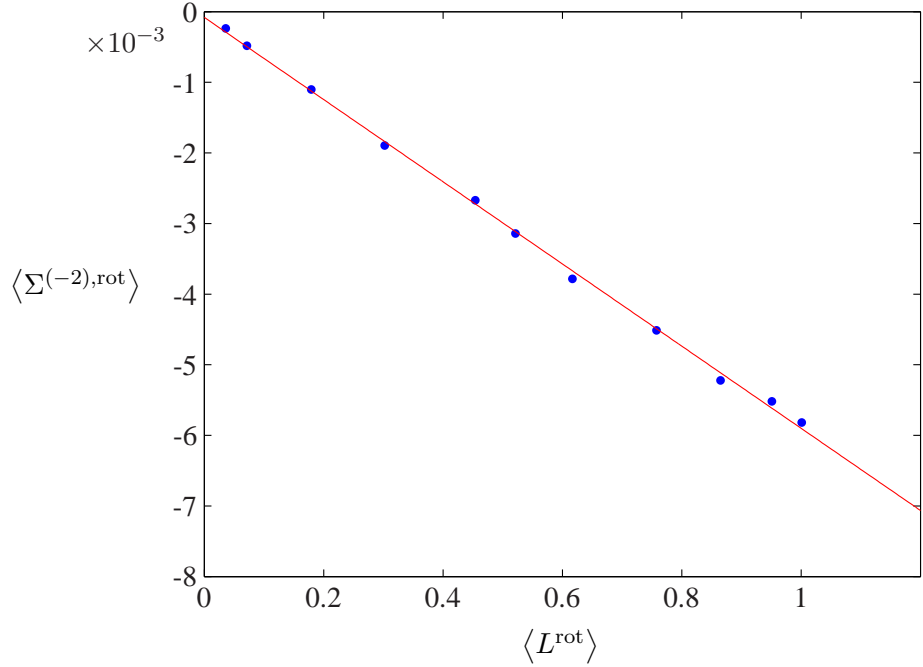}
\caption{\label{fig:propLambdaMinusZwei43} 
The expectation values of $\Sigma^{(-2),\rm rot}$
as functions of $\langle L^{\rm rot}\rangle$ on a $4^3\times 3$ lattice.}
\end{figure}

We have repeated our calculations for the spectral sum $\Sigma^{(-2),\rm rot}$.
The corresponding results for the expectation values in 
Fig. \ref{fig:propLambdaMinusZwei43} show again a linear
relation between the expectation values of this spectral sum and the
traced Polyakov loop. 
\\
This time a linear fit yields  
$\langle\Sigma^{(-2),\rm rot}\rangle=-0. 00582\cdot\langle L^\mathrm{rot}\rangle-8.035\cdot 10^{-5}.$
\section{Exponential spectral sums}\label{exponential}
After the convincing results for sums of inverse powers
of the eigenvalues we analyze the partial exponential spectral sums 
\begin{eqnarray}
\cE_n=\frac{1}{\kappa}\sum_k \bar z_k\sum_{p=1}^{n}
e^{-\lamzk_p}&\Longrightarrow& \cE\equiv \cE_{\rm dim \cD}=\frac{1}{\kappa}\sum_k
\bar z_k \,\tr\exp\left(-\dizk \right)\label{en1a}\\
\cG_n=\frac{1}{\kappa}\sum_k \bar z_k\sum_{p=1}^{n}
e^{-\vert\lamzk_p\vert^2}&\Longrightarrow&
\cG=\cG_{\rm dim \cD}=\frac{1}{\kappa}\sum_k
\bar z_k \,\tr \exp\left(-\dizk^\dagger\, \dizk \right)\label{en1b}
\end{eqnarray}
In particular the last expression is used in a heat kernel regularization
of the fermionic determinant. $\kappa\cG$ has a well-behaved continuum limit if
we enclose the system in a box with finite volume.
We computed the partial sums $\cG_n$ for the flat
connections and various values 
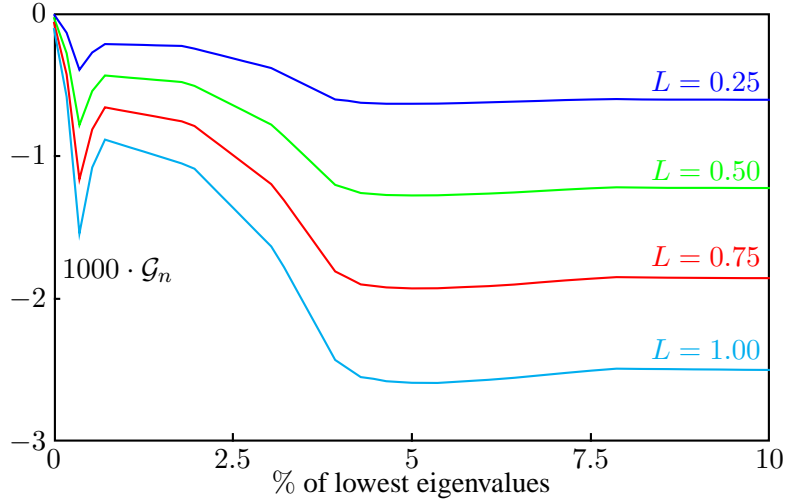
\begin{figure}[ht]
\psset{xunit=1.7mm,yunit=1.9cm}
\begin{pspicture}(-10,.1)(60,-3.2)
\psframe(1,0)(57,-3)
\small
\rput(6,-1.8){$1000\cdot \cG_n$}
\rput(29,-3.3){$\%$ of lowest eigenvalues}
\rput(1,-3.12){$0$}
\rput(57,-3.12){$10$}
\psline(29,-3)(29,-2.94)\rput(29,-3.12){$5$}
\psline(15,-3)(15,-2.94)\rput(15,-3.12){$2.5$}
\psline(43,-3)(43,-2.94)\rput(43,-3.12){$7.5$}
\psline(1,-1)(1.5,-1)\rput[r](.4,-1){$-1$}
\psline(1,-2)(1.5,-2)\rput[r](.4,-2){$-2$}
\rput[r](.4,0){$0$}\rput[r](.4,-3){$-3$}
\rput(52,-.48){\blue $L=0.25$}
\psline[linecolor=blue]
(1,-0.007)(2,-0.137)(3,-0.396)(4,-0.275)(5,-0.216)(6,-0.218)(7,-0.220)
(8,-0.222)(9,-0.225)(10,-0.227)(11,-0.229)(12,-0.248)(13,-0.271)(14,-0.293)
(15,-0.316)(16,-0.339)(17,-0.361)(18,-0.384)(19,-0.428)(20,-0.472)(21,-0.516)
(22,-0.560)(23,-0.604)(24,-0.613)(25,-0.627)(27,-0.634)(29,-0.634)
(31,-0.633)(33,-0.629)(35,-0.624)(37,-0.619)(39,-0.614)(41,-0.608)
(43,-0.604)(45,-0.601)(47,-0.604)(49,-0.605)(51,-0.605)(53,-0.605)(55,-0.606)
(57,-0.606)
\rput(52,-1.1){\green $L=0.50$}
\psline[linecolor=green]
(1,-0.028)(2,-0.280)(3,-0.783)(4,-0.545)(5,-0.436)(6,-0.444)(7,-0.452)
(8,-0.459)(9,-0.467)(10,-0.474)(11,-0.482)(12,-0.509)(13,-0.554)(14,-0.599)
(15,-0.644)(16,-0.689)(17,-0.735)(18,-0.780)(19,-0.861)(20,-0.946)(21,-1.031)
(22,-1.117)(23,-1.202)(24,-1.230)(25,-1.259)(27,-1.273)
(29,-1.276)(31,-1.275)(33,-1.269)(35,-1.263)(37,-1.255)(39,-1.245)(41,-1.235)
(43,-1.226)(45,-1.219)(47,-1.221)(49,-1.223)(51,-1.223)(53,-1.223)(55,-1.224)
(57,-1.224)
\rput(52,-1.7){\red $L=0.75$}
\psline[linecolor=red]
(1,-0.061)(2,-0.429)(3,-1.163)(4,-0.814)(5,-0.659)(6,-0.676)(7,-0.692)
(8,-0.708)(9,-0.724)(10,-0.741)(11,-0.757)(12,-0.790)(13,-0.858)(14,-0.925)
(15,-0.993)(16,-1.061)(17,-1.129)(18,-1.197)(19,-1.310)(20,-1.435)(21,-1.559)
(22,-1.684)(23,-1.808)(24,-1.853)(25,-1.899)(27,-1.920)
(29,-1.927)(31,-1.926)(33,-1.918)(35,-1.911)(37,-1.900)(39,-1.885)(41,-1.871)
(43,-1.859)(45,-1.848)(47,-1.850)(49,-1.852)(51,-1.853)(53,-1.854)(55,-1.855)
(57,-1.855)
\rput(52,-2.35){\cyan $L=1.00$}
\psline[linecolor=cyan]
(1,-0.104)(2,-0.585)(3,-1.542)(4,-1.080)(5,-0.885)(6,-0.913)(7,-0.941)
(8,-0.968)(9,-0.996)(10,-1.024)(11,-1.052)(12,-1.089)(13,-1.180)(14,-1.271)
(15,-1.361)(16,-1.452)(17,-1.543)(18,-1.633)(19,-1.778)(20,-1.941)(21,-2.103)
(22,-2.265)(23,-2.428)(24,-2.488)(25,-2.548)(26,-2.560)(27,-2.577)
(29,-2.588)(31,-2.589)(33,-2.578)(35,-2.567)(37,-2.553)(39,-2.536)(41,-2.519)
(43,-2.503)(45,-2.489)(47,-2.491)(49,-2.492)(51,-2.494)(53,-2.495)(55,-2.497)
(57,-2.498)
\end{pspicture}
\caption{\label{fig:Gaussflat}
The partial Gaussian sums $\cG_n$ for flat connections with different $L$.}
\end{figure}
of the traced Polyakov loop. 
In Fig. \ref{fig:Gaussflat} we plotted those sums for which $10\%$ 
or less of the low lying eigenvalues have been included.
Similarly as for the sums of negative powers of the eigenvalues we conjecture 
that the Gaussian sums $\cG_n$ are good candidates for an order parameter in the
infrared.

\begin{figure}[ht]
\includegraphics{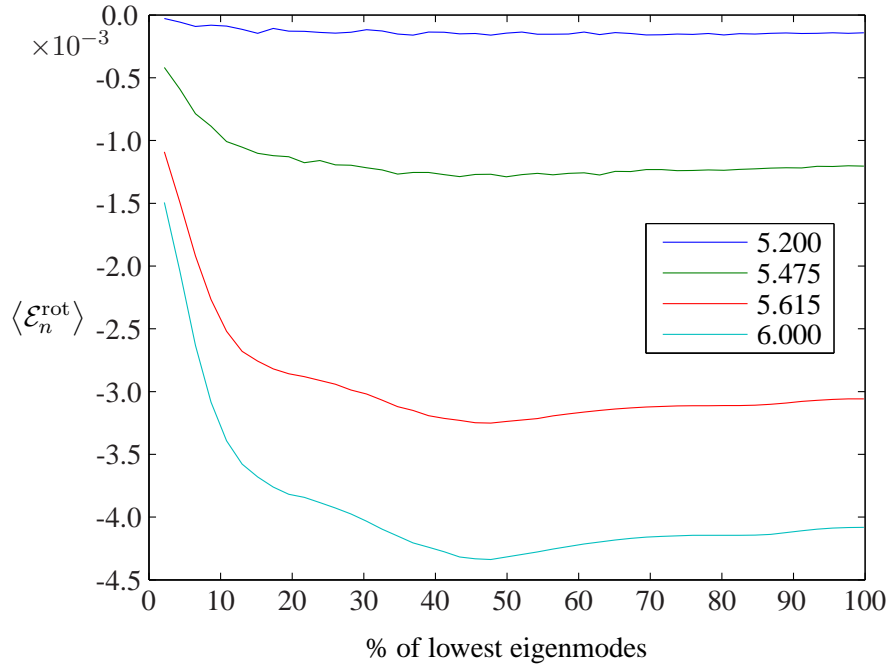}
\caption{\label{fig:expoMinusLambda43} 
Mean exponential sums $\cE_n^{\rm rot}$
on a $4^3\!\times\!3$-lattice near $\beta_{\rm crit}$.
The graphs are labelled with $\beta$.}
\end{figure}

\begin{figure}[ht]
\includegraphics[scale=.95]{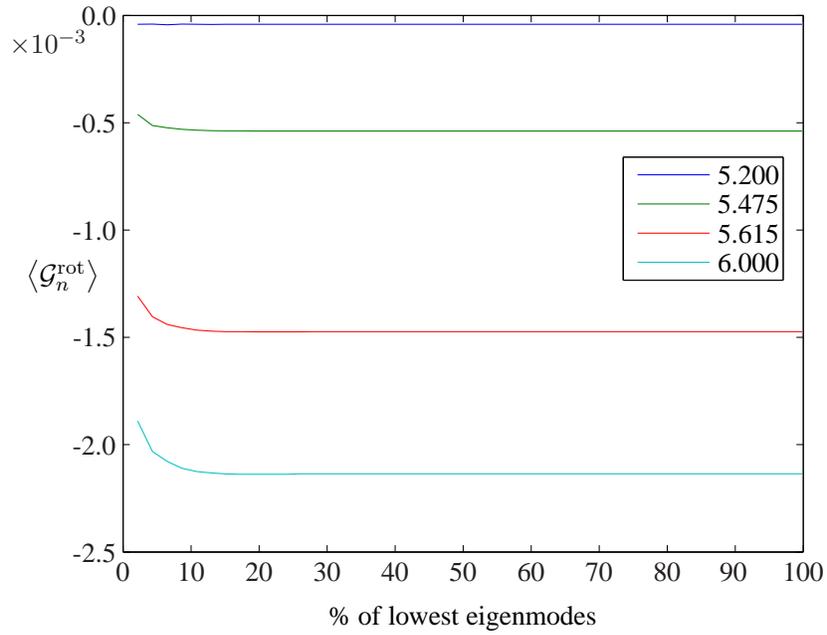}
\caption{\label{fig:expoMinusLaLaQuer43} 
Mean Gaussian sums $\cG^{\rm rot}_n$ on a $4^3\!\times\!3$-lattice near $\beta_{\rm crit}$.
The graphs are labelled with $\beta$.}
\end{figure}

\begin{figure}[h!]
\includegraphics[scale=.95]{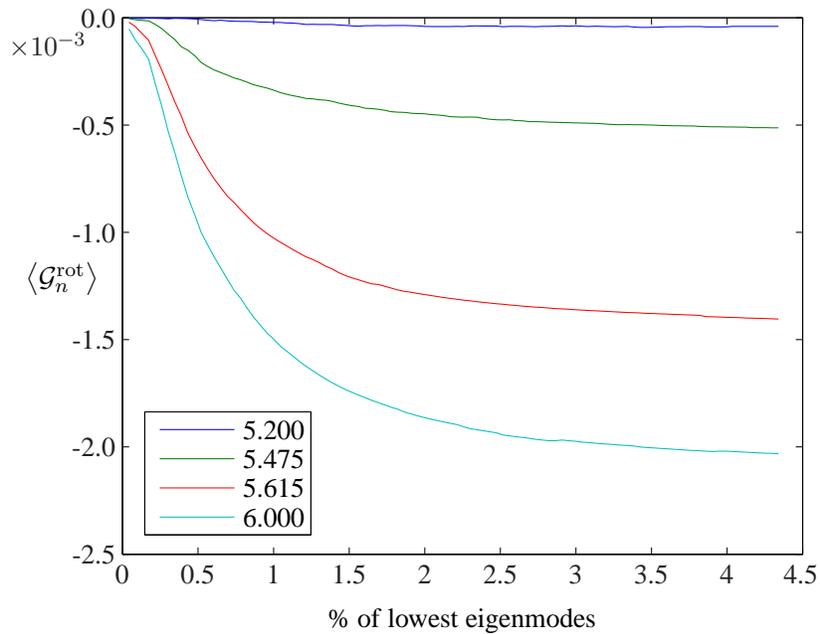}
\caption{\label{fig:expoMinusLaLaQuerAusschnitt43}
Zooming into Gaussian sums $\cG_n^{\rm rot}$
on a $4^3\!\times\!3$-lattice near the phase transition.}
\end{figure}

The expectation values of the partial sums $\cE_n^{\rm rot}$ and $\cG_n^{\rm rot}$ for 
Monte-Carlo generated configurations at four Wilson couplings are plotted 
in Fig. \ref{fig:expoMinusLambda43} and Fig. \ref{fig:expoMinusLaLaQuer43}.
As expected from our studies of flat connections, the Gaussian sums are perfect 
order parameters for the center symmetry. 
They are superior to the other spectral sums considered in this paper,
since their support is even further at the infrared end of the spectrum.
Fig. \ref{fig:expoMinusLaLaQuerAusschnitt43} shows the expectation values 
$\langle\cG^{\rm rot}_n\rangle$
with only $4.5\%$ or less of the infrared-modes included. 
The result is again in qualitative
agreement with that for flat connections in Fig. \ref{fig:Gaussflat}, although in
the Monte-Carlo data the dips are washed out.

\begin{figure}[ht]
\includegraphics{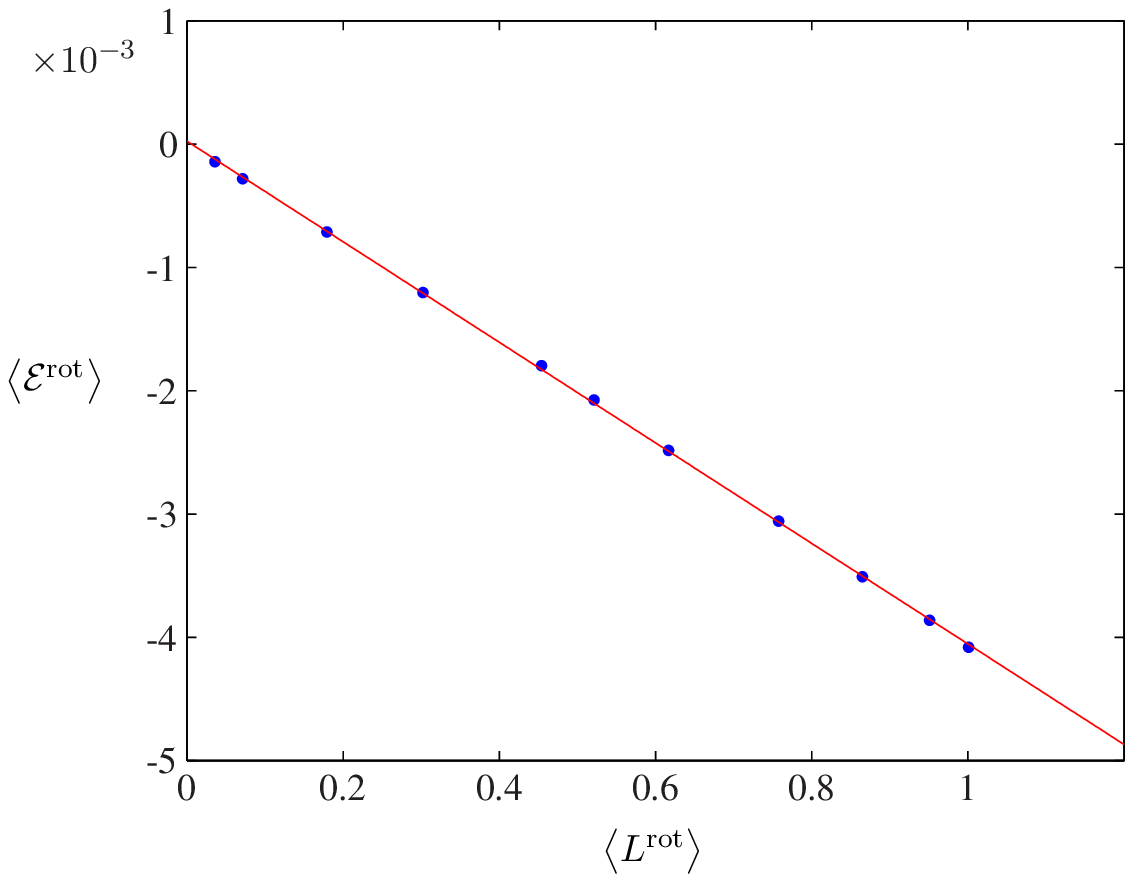}
\caption{\label{fig:propExpoMinusLambda43}
The expectation value of $\cE^{\rm rot}$ as function of $\langle L^{\rm rot}\rangle$
on a $4^3\times 3$ lattice.}
\end{figure}

The Monte-Carlo results for the expectation values $\langle \cE^{\rm rot}\rangle$ and 
$\langle L^{\rm rot}\rangle$ with Wilson couplings in 
table \ref{table:loopsOverBeta} are depicted in Fig. \ref{fig:propExpoMinusLambda43}.
The quality of the linear fit 
\eqnl{
\langle\cE^{\rm rot}\rangle=-0.00408\cdot \langle L^\mathrm{rot}\rangle +2.346\cdot 10^{-5}
\quad ({\rm rmse} = 1.82\cdot 10^{-5}),}{en11}
is as good as for the spectral sum $\Sigma^{(-1)}$.
\\
To estimate the slope and in particular its dependence on the
lattice size we expand the exponentials in $\cE_n$ which results in
\eqnl{
\cE_n=(-)^{\Nt}\sum_{p=0}^\infty\frac{(-1)^p}{(\Nt+p)!}\,\Sigma^{(\Nt+p)}_n.}
{en3}
Since $\Sigma^{(\ell)}$ is proportional to the traced Polyakov loop
for $\ell \leq 3\Nt$ we conclude that $\cE=\cE_{{\rm dim}\cD}$ should 
be proportional to $L$. We can estimate the 
proportionality factor as follows: in the Wilson loop
expansion of $\tr\,\cD^{(\Nt+p)}$ only loops winding around the 
periodic time direction contribute. If we neglect fat loops and only count the straight 
loops winding once around the periodic time direction, then there are
\eqnl{
(m+d)^p\cdot {\Nt+p \choose p}}{en5}
such loops contributing. Actually, for $p\geq \Nt$ there are loops 
winding several times around the time direction. But these have relatively
small entropy and do not contribute much. Hence, with \refs{en3} we arrive
at the estimate 
\eqnl{
\cE\approx (-1)^{\Nt}\sum_{p=0}\frac{(-1)^p}{(\Nt+p)!}(m+d)^p\cdot {\Nt+p
\choose p}\cdot L =\frac{(-1)^{\Nt}}{\Nt!} e^{-(m+d)}L.
}{en7}
In $4$ dimensions and for $m=0$ we obtain the approximate linear relation
\eqnl{
\Nt!\,\cE \approx  (-1)^{\Nt}\cdot 0.0183\cdot L.}{en9} 
For the linear fit \refs{en11} to the MC-data the slope is $3!\cdot 0.00408=0.0245$ instead of $0.0183$.

\begin{figure}[ht]
\includegraphics{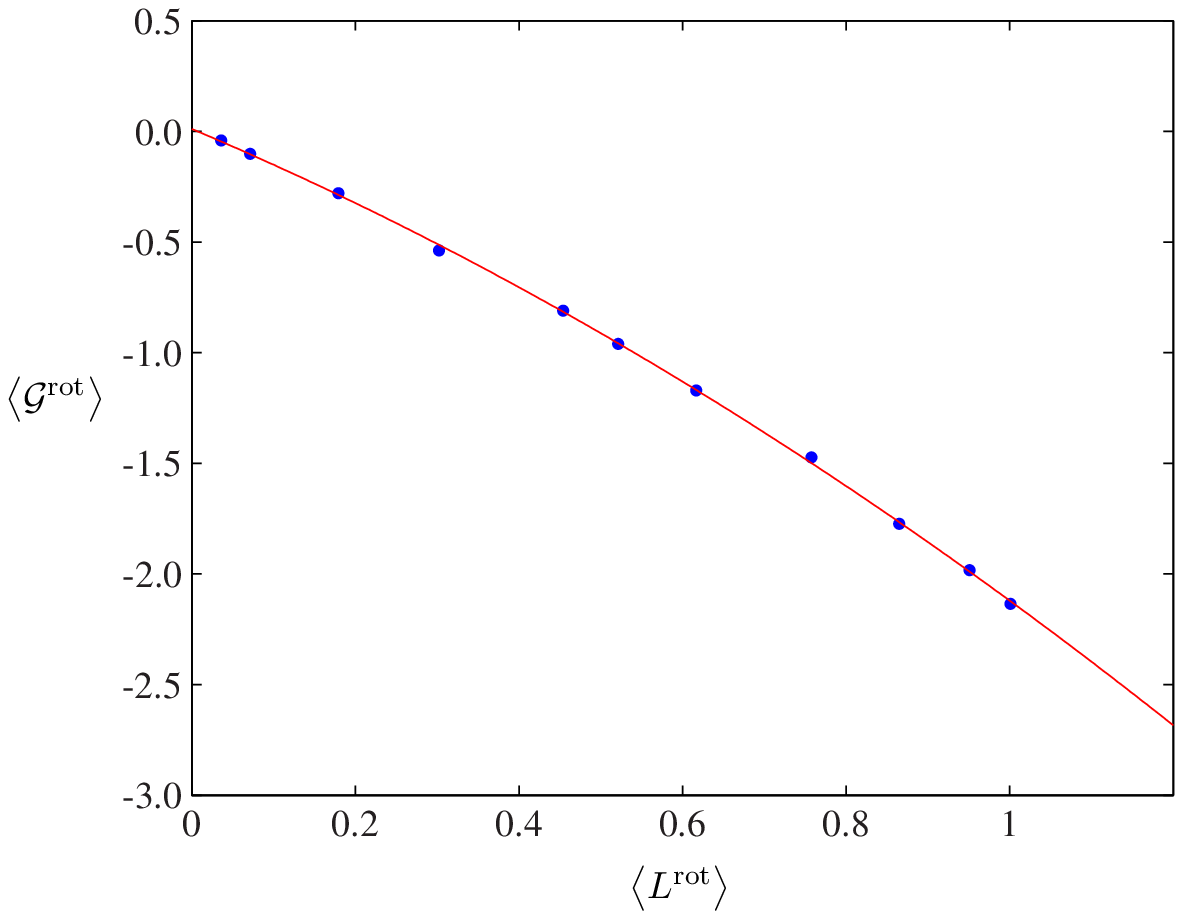}
\caption{\label{fig:quadratExpoMinusLaLaQuer43.eps}
The expectation value of $\cG^{\rm rot}$ as function of $\langle L^{\rm rot}\rangle$
on a $4^3\times 3$ lattice.}
\end{figure}

The Monte-Carlo results for the order parameters $\langle\cG^{\rm rot}\rangle$
and $\langle L^{\rm rot}\rangle$ with Wilson couplings from table \ref{table:loopsOverBeta}
are shown in Fig. \ref{fig:quadratExpoMinusLaLaQuer43.eps}.
In this case the functional dependence is more accurately described by a 
quadratic function,
\eqnl{
\langle\cG^{\rm rot}\rangle=-0.000571 \cdot\langle L^\mathrm{rot}\rangle^2
-0.00156\cdot\langle L^\mathrm{rot}\rangle
+ 1.061\cdot 10^{-5}
\quad ({\rm rmse} =1.453\cdot 10^{-5}),}{en13} 
and this relation is very precise. Since in addition
$\langle \cG_n^{\rm rot}\rangle \approx \langle\cG^{\rm rot}\rangle$
already for small $n$ we can reconstruct the
order parameter $\langle L^{\rm rot}\rangle$
from the low lying eigenvalues of the Wilson-Dirac operator.

\begin{figure}[ht]
\includegraphics{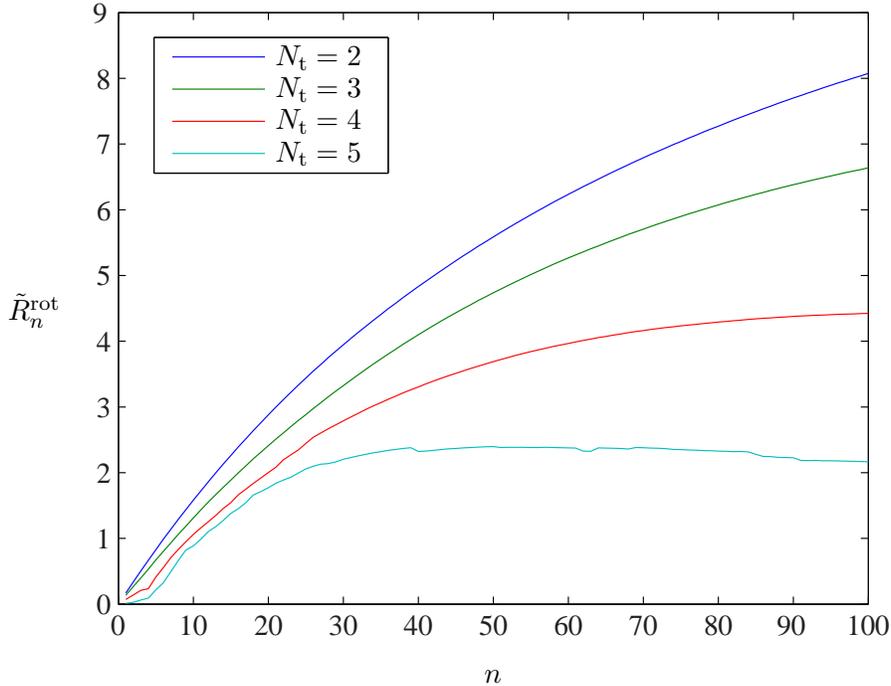}
\caption{\label{fig:expoMinusLaLaQuer6N}
The rations $\tilde R^{\rm rot}_n$ as function of the number $n$
of IR-eigenvalues included. $100$ eigenvalues corresponds to 
approximately  $1\%$ of all eigenvalues.}
\end{figure}

\paragraph{Scaling with $\Nt$:}  
On page \pageref{Finite} we discussed the finite (spatial) size scaling of the 
MC expectation values $\langle\Sigma_n^{\rm rot}\rangle$. We showed that
they converge rapidly to their infinite-$\Ns$ limit, see Fig. \ref{fig:scalingCurves}.
Here we study how the Gaussian sums $\cG_n$ depend on the temporal
extend of the lattice. To that end we performed simulations on
larger lattices with fixed $\Ns=6$, variable $\Nt=2,3,4,5$ and
Wilson coupling $\beta=6.5$. We calculated the ratios
\eqnl{
\tilde R^{\rm rot}_n=\frac{\kappa}{\langle L^{\rm rot}\rangle}\,\langle \cG_n^{\rm rot}\rangle,}{en15}
where we multiplied with the extensive factor $\kappa$ in \refs{gattr1} 
since in the partial sums
\eqnl{
\tilde{\mathcal{G}}_n=\kappa \cG_n=\sum_k \bar z_k\sum_{p=1}^{n}
e^{-\vert\lamzk_p\vert^2}\,,\qquad \vert\lam_p\vert\leq \vert \lam_{p+1}\vert.
}{en17}  
only a tiny fraction of the $5184$ to $12\,960$ eigenvalues
have been included. The order parameter $\langle L^{\rm rot}\rangle$
for the lattices with $\Nt=2,\,3,\,4,\,5$ is $1.9474,\, 1.40194, \,0.932245,\, 0.523142$.
In Fig. \ref{fig:expoMinusLaLaQuer6N} we plotted the ratios $\tilde R^{\rm rot}_n$
for $n$ from $1$ up to $100$.
Note that on the $6^3\times 5$-lattice $n=100$ means less than $0.8\%$
of all $12\,960$ eigenvalues.
This figure very much supports our earlier statements about the quality of the
order parameters $\langle \cG_n^{\rm rot}\rangle$ or $\langle\tilde\cG_n^{\rm rot}\rangle$.
\section{Conclusions}\label{conclusion}
In this paper we studied spectral sums of the type
\eqnl{
\cS_n(f)=\frac{1}{\kappa}\sum_k \bar z_k \sum_{p=1}^n
f\left(\lamzk_p\right)
\mtxt{and}
\hat\cS_n(f)=\frac{1}{\kappa}\sum_k \bar z_k \sum_{p=1}^n
f\left(\vert\lamzk_p\vert^2\right).
}{con1}
where $\{\lamzk_p\}$ is the set eigenvalues of the Wilson-Dirac operator
with twisted gauge field.
Summing over all dim$(\cD)$ eigenvalues the sums over $p$ become traces 
such that 
\eqnl{
\cS(f)=
\frac{1}{\kappa}\sum_k \bar z_k \,\tr f\left(\dizk\right)\mtxt{and}
\hat\cS(f)
=\frac{1}{\kappa}\sum_k \bar z_k \,\tr f\left(\dizk^\dagger\,\dizk\right).
}{con3}
For $f(\lam)=\lam^{\Nt}$ one finds the spectral sum $\Sigma$ which
reproduces the traced Polyakov loop \cite{Gattringer:2006ci}.
Unfortunately this lattice-result is probably of no use in the continuum
limit. Thus we have used functions $f(\lam)$ which vanish
for large (absolute) values of $\lam$. The corresponding sums are
order parameters which get their main contribution from the infrared end 
of the spectrum. Of all spectral
sums considered here, the Gaussian sums $\cG_n$ in \refs{en1b} define
the most efficient order parameters. Besides the $\cG_n$ the spectral sums of
inverse powers of eigenvalues are quite useful as well.
This observation may be of interest since these sums relate 
to the Banks-Casher relation.

It remains to investigate the continuum limits of the spectral
sums considered in this paper. The properly
normalized $\cG_n$ should have a well-behaved continuum limit.
With regard to the conjectured relation between confinement and chiral
symmetry breaking it would be more interesting to see whether
the suitably normalized sums $\Sigma^{(-1)}$ or/and 
$\Sigma^{(-2)}$ can be defined in the continuum theory.
Clearly, the answer to this interesting question depends on the dimension
of spacetime.

\textbf{Acknowledgments:} We thank Georg Bergner, Falk Bruckmann, Christof Gattringer,
Tobias K{\"a}stner and Sebastian Uhlmann for interesting discussions. This project has 
been supported by the DFG, grant Wi 777/8-2.

\end{document}